\documentclass[aps,prd,preprintnumbers,nofootinbib,twocolumn,showpacs,showkeys]{revtex4}
\usepackage{bm}
\usepackage{latexsym}
\usepackage{dcolumn}
\usepackage{amsmath,amsfonts,amssymb}
\usepackage{amsthm}

\begin{document}

\title{Absence of an Effective Horizon for Black Holes in Gravity's Rainbow}

\author{Ahmed Farag Ali} \email[email:]{ahmed.ali@fsc.bu.edu.eg; afarag@zewailcity.edu.eg}
\affiliation{Center for Fundamental Physics,\\ Zewail City of Science and Technology,
12588, Giza, Egypt\\ and \\
Deptartment of Physics, Faculty of Science,\\ Benha University,
Benha, 13518, Egypt}

\author{Mir Faizal} \email[email:]{f2mir@uwaterloo.ca}
\affiliation{Department of Physics and Astronomy, \\  University of Waterloo, Waterloo,
Ontario, N2L 3G1, Canada}

\author{Barun Majumder} \email[email:]{barunbasanta@iitgn.ac.in}
\affiliation{Department of Physics, IIT Gandhinagar, Ahmedabad, India\\ and \\
Deptartment of Physics, Montana State University, Bozeman, MT, USA}
\date{\today}

\begin{abstract}
	We argue that the divergence in time for the asymptotic observer occurs because of specifying the position of the Horizon
	beyond the Planck scale. In fact, a similar divergence in time will also occur for an in-going observer
in Gravity's Rainbow, if we again specify the  position of the Horizon beyond the Planck scale.
On the other hand, if we accept the occurrence of a minimum measurable length scale associated with a universal invariant maximum
energy scale in Gravity's Rainbow, then the time taken by both the in-going and asymptotic observers will be finite.
\end{abstract}

\pacs{04.70.Dy, 04.60.Bc}
\keywords{Black Holes, Horizon, DSR, MDR, Gravity's Rainbow}

\maketitle

%
%

The loss of information in the black hole   has been an important problem in quantum gravity \cite{1}. The black hole complementarity has been an attempt to revolve this information loss problem \cite{a}. According to black hole complementarity, an in-going observer will observe classical metric and cross the Horizon in finite amount of time. However, the time taken for the in-going observer will diverge for an
asymptotic observer and he will observe information in form of Hawking radiation.
There is an active debate about the information that an asymptotic observer can obtain  from the Hawking radiation and an
alternative point of view called the Firewall proposal has also been recently proposed \cite{b}.
\par
All these approaches
do not take into account the modification to the theory that will occur due to Planck scale phenomena. This might be because
it is not possible to  directly analyze Planck scale effects.
However, it is possible to analyze these
effects from an effective field theory perspective.
Thus, there will be a remnant behavior of these Planck scale phenomena for black hole physics at a lower energy scales. In this paper,
we will analyze the
time taken for an in-going observer to cross the Horizon,  as measured by the asymptotic observer and the in-going observer, using the
remnants of Planck scale effects.
All most all models of quantum gravity predict that it would not be possible to define spacetime below Planck scale
\cite{d}-\cite{d1}.
The existence of minimum length will
 deform the dispersion relation in special relativity, and the resultant theory is called  the doubly special relativity theory
 \cite{2}-\cite{3}. A generalization of the doubly special relativity to include general relativity, gives rise to  Gravity's
 Rainbow \cite{n1}-\cite{n2}. In this paper, we will demonstrate that in Gravity's Rainbow, the time taken by both the in-going
 observer and the asymptotic observer will be finite, if the measurements are made to the level of accuracy allowed in the Gravity's Rainbow.
We will argue that the  divergence of time for the asymptotic observer occurs because of specifying the location of Horizon
 beyond Planck scale. In fact, we will demonstrate that in Gravity's Rainbow, a divergence in time will occur even for an in-going observer,
 if we try to specify the location of
 Horizon beyond Planck scale. However, if we take the idea of Gravity's Rainbow seriously, and assume that there is a minimum length scale
 which remains
 invariant for all observers, the time measured by the asymptotic observer will also become finite. Some recent developments in this field
 can be found in \cite{allref}-\cite{temp}.
\par
Now we will explicitly, calculate the time taken by an in-going observer and an asymptotic observer in gravity's Rainbow.
The Schwarzschild metric in gravity's Rainbow can be written as
\begin{eqnarray}
dS^2 &=& \frac{(1-\frac{2  G (E) M}{ r})}{f^2(E)}dt^2 - \frac{dr^2}{(1-\frac{2 G(E) M}{r})g^2(E)} \nonumber \\
&-& \frac{r^2}{g^2(E)}d\theta^2 - \frac{r^2 \sin^2 \theta}{g^2(E)}d\phi^2 ~~,
\end{eqnarray}
where we have chosen units such that $  \hbar=1 $ and $G(E)  = {G^{\prime}(E)}/{c^2}=G^{\prime}(E) {f^2(E)}/{g^2 (E)}$,
where $G^{\prime}(E)$ is the scale dependent Newton's constant and the speed of light being $c=g(E)/f(E)$ in Gravity's Rainbow.
Even though there are various choices  of  phenomenologically motivated  Rainbow functions  \cite{allref}-\cite{temp}, 
all these Rainbow functions  can be   divided into two classes  i.e., 
those Rainbow functions in which it is possible to probe trans-Planckian physics, 
and those Rainbow functions in which it is not possible to probe trans-Planckian physics. 
The Rainbow functions proposed  by Amelino-Camelia {\it et.al} in \cite{amea}, 
to explain hard spectra from gamma-ray burster's at cosmological distances are
given by $ f(E) = (e^{\alpha E/E_p } -1) \times (\alpha E/E_p)^{-1}$ and $ g(E) = 1$. 
Hence, they form an example of those Rainbow functions in which it is possible to probe trans-Planckian physics, 
if $\alpha = 1$. 
The Rainbow functions in which it is not possible to probe trans-Planckian physics can be sub-divided into those Rainbow 
functions for which $g(E) /f(E) \to 0$ as $E \to E_p$,  and  those for which 
$g(E) /f(E)  \neq 0$ as $E \to E_p$. 
In the later case both $f(E)$ and  $g(E)$ can diverge individually but the limit $g(E) /f(E)  \neq 0$ as $E \to E_p$, 
holds. However, their  divergence restricts the analysis of  trans-Planckian physics.
The Rainbow functions  proposed by 
Magueijo and  Smolin consistent with a constant velocity of light are given by 
$ f(E) = g(E) = ( 1 - \lambda E/E_p)^{-1}$ \cite{2}. For $\lambda = 1$, they 
form an example of those Rainbow functions for which 
$g(E) /f(E)  \neq 0$ as $E \to E_p$, but it is not possible to probe trans-Planckian physics.
Finally, the Rainbow functions proposed by Amelino-Camelia {\it et.al} in \cite{ame}-\cite{ame1}
are given by $f(E)=1$ and $g(E)=\sqrt{1-\eta(E/E_P)^n}$. 
They form an example of those rainbow functions for which 
$g(E) /f(E) \to 0$ as $E \to E_p$, if $\eta = 1$. 
In this paper, we will demonstrate that for this class of Rainbow functions, 
the black hole information paradox can be resolved. These Rainbow functions proposed by Amelino-Camelia {\it et.al} in \cite{ame}-\cite{ame1}, 
are consistent with loop quantum gravity and non-commutative geometry  \cite{2}. Hence,  
we can state the main conclusion of this paper is that the deformation of the energy-momentum  dispersion relation by loop 
quantum gravity or the non-commutative geometry can lead to a resolution of the black hole information paradox.
As non-commutative geometry is closely related to string theory \cite{aq}, it can be stated that deformation of the energy-momentum  dispersion
because of stringy effects can lead to a maximum energy scale in the low energy effective theory and which in turn will lead to a resolution 
of black hole information paradox. In fact in string theory, it is not possible to probe spacetime below string length scale \cite{string}.  
So, string theory comes naturally equipped with a minimum length scale, and this also corresponds to a maximum energy scale, which  
 in turn restricts the analysis of trans-Planckian physics. 
It may be noted that our analysis is based on the existence of a maximum energy scale which restricts the analysis of 
trans-Planckian physics, and so, our analysis is not valid for those choices of Rainbow functions in which it is possible 
to probe trans-Planckian physics.  
Now 
we choose the  Rainbow functions for our analysis to have the 
particular form, $f(E)=1$ and $g(E)=\sqrt{1- (E/E_P)^n}$ .   We continue to write $f(E)$ and $g(E)$ throughout
this paper, but we refer to this choice. This choice of Rainbow functions was also initially proposed by Amelino-Camelia {\it et.al}
in \cite{ame}-\cite{ame1},
to show that certain data analysis provides sensitivity to effects introduced genuinely at the Planck scale.  It may be noted that in the
limit $E \to E_P$, we obtain $g(E) \to 0$, and in the limit $E \to 0$, we obtain $g(E) \to 1$, {which may be considered as the asymptotic
limit when we get back general relativity. } In the original formalism of Gravity's Rainbow by Magueijo and Smolin \cite{n1},
Newton's constant did not dependent on energy, and all the energy dependence was absorbed in to the Rainbow functions. However,
there is no reason to limit the energy dependence in this way, and as there is a choices in choosing  phenomenologically motivated
Rainbow functions  \cite{allref}-\cite{temp}, it is possible to
new Rainbow functions in such a way that the Newton's constant is taken to be energy dependent. We will perform our analysis using
this formalism
of Gravity's Rainbow, where the Newton's constant is energy dependent.

{Now if the energy $E_p$ is required to probe spacetime at a length scale $L_P$, then in Gravity's Rainbow we can not probe spacetime
 below this length scale. This is because we will require a energy greater than $E_p$ to do that, and that is
 not allowed in Gravity's Rainbow. We can only specify measurement of distances to an accuracy of $L_p$.
 So, we have to limit the resolution to which a distance can be measured. Thus, we  introduce  an expression
 to indicate the resolution to which an quantity is measured. So, for a quantity $Q$,  the expression
 $\langle Q \rangle = Q_p$,  indicates that the quantity $Q$ is measured to the accuracy $Q_p$.
 In other words, the minimum measurable value of $Q$ is $Q_p$ and not zero.
 For example, in Gravity's Rainbow $\langle 1/ E \rangle = 1/ E_p$, as there is a maximum value for the energy
 in Gravity's Rainbow, there is also a minimum value for $1/E$ in Gravity's Rainbow.
 However, we can now express  this in terms of a minimum distance $L$,  or the scale at which spacetime will be probed by
 an particle of energy $E$. So, we can write for any particle  with energy $E$, the  minimum distance  or the scale at which spacetime
 will be probed is given by,
 $\langle L \rangle = L_p$.  As this holds for any position in spacetime, it also holds for the Horizon. So,
 we can only specify the position of the Horizon with the accuracy,  $\langle r - 2 G(E)M\rangle >
 L_p$.  If we try to  specify the position for the Horizon with greater accuracy, we would have to
 probe distances smaller than $\langle r - 2 G(E)M\rangle >
 L_p$, but this would require an energy greater than $ E_p$, which is not allowed in Gravity's Rainbow.
 In fact, in Gravity's Rainbow  the
 manifold structure of spacetime breaks down at $ E = E_p$, due to the divergence of the Rainbow functions.
 Hence, we can only specify the position of the Horizon with the accuracy $\langle r - 2 G(E)M\rangle >
 L_p$.

 It may be noted that the
 divergence in time  for an asymptotic observer
 occurs when we measure the spacetime with arbitrary  accuracy, i.e., $\langle r - 2 G(E)M\rangle = 0$,
 but this can only be done if there is no maximum energy scale in the theory. The existence of a maximum energy scale prevents  the measurement
 of spacetime with arbitrary  accuracy, such that $\langle r - 2 G(E)M\rangle >
 L_p$.
 We will now demonstrate that  limiting the measurement of $\langle r - 2 G(E)M\rangle >
 L_p$ will make the  time for asymptotic observe finite. On the other hand if we measure
 distance with the accuracy, $\langle r - 2 G(E)M\rangle =
 L_p$, the time for both the  asymptotic and the in-going observers will diverge due to  $g(E) \to 0$ as $E \to E_p$,
 before it can diverge for the asymptotic observer at $\langle r - 2 G(E)M\rangle =0$,
 due to effects coming from general relativity.
 So, the main finding of this analysis is that the divergence of time for the asymptotic observer occurs
 because of specifying the position of Horizon with arbitrary accuracy, and this is not allowed
 in Gravity's Rainbow. Hence, the time for an  asymptotic observer will also be finite in Gravity's Rainbow. }

\par
For an   incoming particle radially falling into the black hole  with velocity $v^{\mu}= dx^{\mu}/{ds}$, we have $v^2 = v^3 =0$, due to the
radial nature of its motion.  The geodesic equation in Gravity's Rainbow can be  written as
\begin{equation}
\frac{dv^0}{ds} = - \Gamma^0_{\mu \nu}(E)~v^{\mu}v^{\nu} ~~.
\end{equation}
This equation can be integrated to give
\begin{equation}
v^0 = \frac{k f^2(E)}{(1-\frac{2 G(E)M}{r})} ~~,
\end{equation}
where $k$ is a constant. We also have the normalization condition
$
g_{\mu \nu}v^{\mu}v^{\nu}=1 $.
Now we can easily calculate the radial component of the velocity as
\begin{equation}
v^1 = -g \left(k^2 f^2(E) - 1 + \frac{2 G(E) M}{r}\right)^{1/2}
\end{equation}
As the particle is falling in so $v^1$ is negative,  and
\begin{eqnarray}
\frac{dt}{dr} = \frac{v^0}{v^1} &=& -\frac{kf^2(E)}{g(E)} \left(1-\frac{2 G(E) M}{r}\right)^{-1} \nonumber \\
&\times &  \left(k^2 f^2(E) -1 + \frac{2 G(E) M}{r}\right)^{-1/2}.
\end{eqnarray}
{Thus, we get an expression for time for an asymptotic observer,
\begin{equation}
t = -\frac{2M f(E)}{g(E)} \log(r-2 G(E)M) - \frac{(2k^2 f^2(E)+1)r}{2f(E)g(E)k^2} + C~~.
\end{equation}
where $C$ is a finite integration constant.
This time diverges at $ r - 2 G(E)M =0$,  however, we cannot make measurements with such an accuracy.
This is because this
 would imply,
$\langle r - 2G(E)M \rangle =0$, and it will take energy greater than $E_p$ to make   measurements with such an accuracy.
So, it is not possible to specify a distance with the accuracy
more than the Planck length in Gravity's Rainbow, because  an  energy greater than Planck energy is required to do that, and
this would make the Rainbow functions vanish, breaking the spacetime description all together. In fact, a central assumption
in doubly special relativity and hence in Gravity's Rainbow is that there is an invariant energy scale which remains same for all the observers.
So, in Gravity's Rainbow we can not measure spacetime with arbitrary precision,
as it would require an energy greater than $E_p$ to do that.   We can only put
$\langle r - 2G(E)M \rangle > L_p$.
Now if $\epsilon$ is any suitable small parameter with dimensions of length, we can write
$\langle r - 2G(E)M \rangle =  L_p + \epsilon $. It may be noted that in the limit,
$\epsilon \to 0$, we have $\langle r - 2G(E)M \rangle = L_p$. It will require an energy
$E_p$ to probe spacetime at this scale, and this will cause $g(E)$ to diverge, leading to the divergence of time for
the asymptotic observer. However, for $\langle r - 2G(E)M \rangle > L_p$, and hence, $\epsilon \neq 0$,  we have
\begin{equation}
t = -
\frac{2M f(E)}{g(E)} \log(L_p + \epsilon ) - \frac{(2k^2 f^2(E)+1)r}{2f(E)g(E)k^2} +
C
~~.
\end{equation}
Thus, we obtain a finite time for the in-going observer.
 The particle takes finite time to reach the event horizon even when measured by an asymptotic observer. It may be noted that the time diverges
 due to Rainbow functions,
 $t \to \infty$, as $\langle r -2 G(E) M \rangle = L_p $ because of $g (E) \to 0$,
 much it diverges due to effects coming from general relativity which occur at $\langle r - 2 G(E) M\rangle =0$. In fact,   in Gravity's Rainbow
 it is  not even allowed to take   $\langle r - 2 G(E) M\rangle =0$,
 as it contradicts the existence of maximum energy. Thus, the divergence of time   occurs, if the position of Horizon is
 specified with an accuracy beyond that which is allowed in Gravity's Rainbow. }

In fact,  it is possible to demonstrate that such a divergence occurs even for an in-going observer at $\langle r - 2 G(E)M\rangle =
 L_p$, and this is an evidence that it occurs due to specifying the position of Horizon beyond Planck scale.
Now we will analyse  a non-static system by making  the following transformation
\begin{eqnarray}
\tau &=&  \frac{t}{f(E)} + m(r), \nonumber \\
\rho &=&\frac{t}{f(E)} + n(r) .
\end{eqnarray}
{Here we follow the method as adopted in \cite{dirac},  and choose suitable functions $m$ and $n$, such that
\begin{eqnarray}
 m^{\prime} = \frac{2 G(E) M}{r} n^{\prime}, \nonumber \\  {m'}^{2} - \frac{2 G(E) M}{r} {n'}^{2} &=& -\frac{1}{(1-2 G(E)M /r) g^2(E)},  \label{ch1}
\end{eqnarray}
where prime means derivative with respect to $r$.
Now we obtain  }
\begin{eqnarray}
d\tau^2 - \frac{2 G(E) M}{r} d \rho^2 &=& \frac{(1-\frac{2 G(E)M}{r})}{f^2(E)}dt^2 \nonumber \\
&-& \frac{dr^2}{(1-\frac{2 G(E) M}{r})g^2(E)}  ,
\end{eqnarray}
  Solving  this equation  for $n'$ and $m'$,  we obtain
\begin{equation}
n' - m' = \frac{1}{g(E)} \sqrt{\frac{r}{2 G(E)M}} ~~.
\end{equation}
So, we obtain our radial coordinate as
\begin{equation}
r = \xi g^{2/3}(E) (\rho - \tau)^{2/3} ~~,
\end{equation}
where $\xi = (3\sqrt{2 G(E)M}/2)^{2/3}$, and the metric can be written as
\begin{eqnarray}
ds^2 &=& d\tau^2 - \frac{2 G(E)M}{\xi g^{2/3}(E)(\rho - \tau)^{2/3}} d\rho^2 \nonumber \\
&-& \frac{\xi^2 (\rho - \tau)^{4/3}}{g^{2/3}(E)}(d\theta^2 + \sin^2 \theta d\phi^2) ~~.
\end{eqnarray}
{Thus, we obtain
\begin{equation}
 (\rho - \tau)=\frac{4 G(E)M}{3g(E)}.
\end{equation}
This is finite in general relativity  \cite{dirac}, if we neglect the effect of the Rainbow functions.
However,  in Gravity's Rainbow  we have to again limit the accuracy with which the position of the Horizon is specified to
 $\langle r-2 G(E)M\rangle  > L_p$. This is because,   we will require an particle with energy
 $E_p$ to specify the position of Horizon with the accuracy,
 $\langle r - 2 G(E) M\rangle  = L_p$. However, in the limit $E \to E_p$, the Rainbow function $g(E)\to 0$,
 and thus $(\rho - \tau)={4 G(E)M
}/{3g(E)} \to \infty$.
So, the time even diverges for an in-going observer, if the position of the Horizon is specified with accuracy
$\langle r - 2 G(E) M\rangle  = L_p$. }
Thus, in Gravity's Rainbow, the time taken by the in-going observer would  also diverge, if the position of Horizon is
specified beyond the Planck scale. This fits well with the idea that the divergence in time is actually an signal of breakdown
of spacetime description of quantum theory of gravity,
which occurs because of specifying a point in spacetime beyond the Planck scale.

This result makes more sense in the light of effective field theories. Both general relativity and Gravity's Rainbow are effective theories of
some true theory of quantum gravity. Now usually there are no special points in the geometry of spacetime,
and hence, the Gravity's Rainbow would agree with general relativity.
But in case of a black hole, as the Horizon is a set of very specific point
in spacetime, Gravity's Rainbow predicts that
 spacetime description breaks down at this scale, and hence it is not allowed
 to measure the point in spacetime  with such an accuracy. So, effectively in Gravity's Rainbow,
 the in-going particle takes only a finite time to cross the Horizon.
In fact, a similar effect even occurs
for an in-going observer  in Gravity's Rainbow, and the time measured by
an in-going observer  also diverges, if  the exact position of the Horizon is specified.
Thus, the divergence of time is an general feature which occurs because of specifying
the Horizon  beyond the Planck scale. On the other hand, if we specify the Horizon, within the limits
in which spacetime description still holds, then the time measured by both the in-going observer
and the asymptotic observer is finite.

This picture becomes more clear in the light of perturbative string theory.
This is because  a measurable  minimum length can
also be inferred from perturbative string theory. However, in perturbative string theory, string scattering amplitudes  are measured on a fixed background spacetime.
The back reaction of the string theory effects on the geometry and topology of the background spacetime,
would bring the discussion out of domain of perturbative string theory into some non-perturbative effects in string theory.
The perturbative string theory also implies the existence of a minimum length scale,
and
this is because even though there is no cut off in geometry of spacetime in perturbative string,
there is an effective cut off in the ability to probe this spacetime  \cite{string,Ali:2009zq,string1}. As in string theory, strings are used
to probe spacetime, so, it is impossible to probe spacetime below the string length scale.  In perturbative
string theory the Horizon does form, however,  we  never know when the string actually cross it, and this makes the time measured by the asymptotic observer finite. It is similar to
the double slit experiment in quantum mechanics. Even though there is only one particle and two slits,  if  we can not know through which slit the
particle pass through, it is assumed to pass through both of them. In the same way, even though there is a Horizon, as we can never know when a string cross it, so effectively, it appears as if there is no Horizon.

Thus, the black hole information paradox can be resolved without any need for black hole complementarity
or Firewall, if the idea of minimum length is taken seriously.  It may be noted that we have not discussed the effect on Hawking radiation from the existence of minimum length. It is both important and interesting to understand what happens to Hawking radiation.
It may also be noted that the surface temperature of a black hole is known to be vanishing in Gravity's Rainbow as the black hole mass reaches a remnant state at which the specific heat  also vanishes \cite{temp}, and this is an indication of the fact that effectively the  Horizon does not form for the black holes. In fact, effectively a Horizon can never  form for  any effective field theory on spacetime generated from Gravity's Rainbow. This is because the Horizon is an exact set of points in spacetime, however,  the metric structure of spacetime breaks down at Planck scale, and so, it is not possible to specify a set of points in spacetime exactly.

\section*{Acknowledgments}

The authors gratefully thank the anonymous referees for enlightening comments and suggestions
which substantially improved the quality of the paper. 
The work of AFA is supported by the CFP at Zewail City of Science Technology and by Benha University (www.bu.edu.eg), Egypt.

\end{document}